\documentclass[a4paper,fleqn,usenatbib]{mnras}

\usepackage{newtxtext,newtxmath}

\usepackage[T1]{fontenc}
\usepackage{ae,aecompl}

\usepackage{graphicx}	
\usepackage{amsmath}	
\usepackage{amssymb}	
\usepackage{color}		
\usepackage[dvipsnames]{xcolor}

\defcitealias{sarah15}{B15}

\newcommand{\cp}{\citep}
\newcommand{\ct}{\citet}
\newcommand{\cpa}{\citepalias}
\newcommand{\cta}{\citetalias}
\newcommand{\HI}{H{\sc{i}}}
\newcommand{\HII}{H{\sc{ii}}}
\newcommand{\HH}{H$_{2}$}
\newcommand{\HA}{H$\alpha$}
\newcommand{\dolphot}{{\sc{dolphot}}}

\title[Star formation in the outskirts of DDO 154]{Star formation in the outskirts of DDO 154: A top-light IMF in a nearly dormant disc}

\author[Watts et al.]{Adam B. Watts,$^{1}$\thanks{E-mail: adam.watts@uwa.edu.au}
Gerhardt R. Meurer,$^{1}$\thanks{E-mail: gerhardt.meurer@uwa.edu.au}
Claudia D. P. Lagos,$^{1}$
Sarah M. Bruzzese$^{1}$,
\newauthor{Pavel Kroupa$^{2,3}$, Tereza Jerabkova$^{2,3,4}$}
\\
$^{1}$International Centre for Radio Astronomy Research, The University of Western Australia, Crawley, WA, Australia\\
$^2$ Helmholtz-Institut für Strahlen- und Kernphysik (HISKP), Universität Bonn, Nussallee 14-16, 53115 Bonn, Germany\\
$^3$ Charles University in Prague, Faculty of Mathematics and Physics, Astronomical Institute, V Hole¨ovickách 2, CZ-180 00 Praha 8, Czech Republic\\
$^4$ European Southern Observatory, Karl-Schwarzschild-Straße 2, 85748 Garching bei München\\
}

\date{Accepted XXX. Received YYY; in original form ZZZ}

\pubyear{2015}

\begin{document}
\label{firstpage}
\pagerange{\pageref{firstpage}--\pageref{lastpage}}
\maketitle

\begin{abstract}
We present optical photometry of Hubble Space Telescope (\emph{HST}) ACS/WFC data of the resolved stellar populations in the outer disc of the dwarf irregular galaxy DDO 154.  The photometry reveals that  young main sequence stars are almost absent from the outermost \HI\ disc. Instead, most are clustered near the main stellar component of the galaxy. We constrain the stellar initial mass function (IMF)  by comparing the  luminosity function of the main sequence stars to simulated stellar populations assuming a constant star formation rate  over the dynamical timescale. The best-fitting IMF is deficient in high mass stars compared to a canonical Kroupa IMF, with a best-fit slope $\alpha = -2.45$ and upper mass limit $M_U = 16\ M_{\sun}$.  This top-light IMF is consistent with predictions of the Integrated Galaxy-wide IMF theory.  Combining the \emph{HST} images with \HI\ data from The \HI\ Nearby Galaxy Survey Treasury (THINGS) we determine the star formation law (SFL) in the outer disc. The fit has a power law exponent  $N = 2.92 \pm0.22$ and zero point $A=4.47 \pm 0.65 \times 10^{-7} \ M_{\sun} \ \text{yr}^{-1} \ \text{kpc}^{-2}$. This is depressed compared to the  Kennicutt-Schmidt Star Formation Law, but consistent with weak star formation observed in diffuse \HI\ environments. Extrapolating the SFL over the outer disc implies that there could be significant star formation occurring that is not detectable in \HA. Last, we determine the Toomre stability parameter $Q$ of the outer disc of DDO 154 using the THINGS \HI\ rotation curve and velocity dispersion map. 72\% of the \HI\ in our field has $Q\leq 4$ and this incorporates 96\% of the observed MS stars. Hence 28\% of the \HI\ in the field is largely dormant.
\end{abstract}

\begin{keywords}
galaxies: individual (DDO 154) -- galaxies: dwarf -- galaxies: stellar content -- stars: massive -- stars: luminosity function
\end{keywords}



\section{Introduction}

The Initial Mass Function (IMF) and the Star Formation Law (SFL), are two empirical relationships that allow us to characterise star formation in galaxies from relatively simple observables.  While they are commonly thought to be universal it is important to test these assumptions in extreme environments. The outer disk of the ISM rich, dark matter dominated dwarf irregular galaxy DDO154 provides us with one such extreme environment. The stellar IMF, originally proposed by \ct{salt55}, gives the number density $N(m)$ of new stars of mass $m$ formed in an event as a single slope power law $N(m) dm \propto m^{\alpha} dm$ with $\alpha = -2.35$. It  remains one of the most enduring star formation relationships, and is used in determinations of total star formation rates (SFR) from various tracers such as \HA\ and ultraviolet emission  \cp[e.g.][]{kennicutt89, kennicutt98, meurer09},  star formation histories (SFH) from colour magnitude diagrams (CMD) of resolved stellar populations \cp[e.g.][]{dolphin02,barker07,hillis16,williams09,williams13}, total stellar mass estimates from luminosities \cp[e.g.][]{baldry08,tortora10,taylor11,moffett16}, and supernova rates \cp[e.g.][]{barris06,mann08,scanna05}. Since Salpeter's original proposal, the IMF has been found to turn over below $1 \  M_{\sun}$ and is commonly parameterised as a broken power law \cp{kroupa01} or log-normal distribution \cp{chabrier03}. Theoretically its shape has been shown to be invariant \cp[e.g.][]{krumholz11,krumholz12} or variable  \cp[e.g.][]{bate03,bate05,bonnell08,motta16} depending on the star-forming conditions,  or whether it is integrated over the galaxy \cp{weidner05}. Many observations of the Milky Way and Magellanic Clouds have shown little to no evidence of variations outside statistical uncertainties \cp[e.g.][]{freedman85,scalo85,phelps93,elmegreen99,bastian10}, which has lead to  the assumption of a universal IMF  \cp{scalo86,kroupa01,chabrier03}. 

 The universality of the IMF has become a hotly debated topic as an increasing number of observational studies have provided evidence of real, detectable variations \cp[e.g.][]{cappellari12,davis17,kalari17,romano17,schneider18}. A non-universal high mass slope ($m> 1 M_{\sun}$, from here just `slope') has been invoked to explain  high mass to light ratios in low mass galaxies, deficiencies of \HA\ emission, and high x-ray emission from ultra compact dwarf galaxies \cp{hoversten08,lee04,lee09,pflamm07, meurer09,dabringhausen10}. These phenomena are thought to be caused by an IMF that varies with star formation intensity,  galactocentric radius, metallicity and local gas density \cp{gunawardhana11,meurer09,marks12}.   There is also evidence for a varying IMF at masses $m <1 M_{\sun}$ in early type galaxies that could correlate with metallicity and velocity dispersion \cp{labarbera15,conroy12,vaughan16,mn15},   and in local group dwarf galaxies \cp[e.g.][]{geha13,kalirai13,gennaro18}. 
 
  In most studies of resolved stellar populations in nearby galaxies it is common to derive the SFH from the CMD by assuming an IMF \cp[e.g.][]{geha98,harris04,harris09,barker07, cignoni12,weisz11,weisz13,weisz15a}, but if the IMF is not universal, rather steeper or deficient in high mass stars, the  SFH could be inferred to be halted or declining when the SFR is actually constant \cp{meurer09}.   Here we extend the method of \ct[][hereafter B15]{sarah15} to use the less common approach of deriving the IMF of a stellar population by assuming a plausible form of the SFH. The IMF is thought to  show variation in extreme environments such as the extended  \HI\ discs of gas rich star-forming galaxies. These regions have low gas and stellar surface densities with correlated diffuse UV emission indicative of low intensity star formation \cp{thilker05, thilker07}, thought to be deficient in high mass stars \cp{elmegreen04}. There are competing models for the deficiency of high mass stars in low density regions: stochasticity of a universal IMF \cp[e.g.][]{koda12}, or a varying IMF \cpa[e.g.][]{sarah15}.
  
One model for a varying IMF is the integrated galaxy IMF \cp[IGIMF, e.g.][]{pflamm091}. Where many theories adopt a model for star formation where stars can form anywhere, the IGIMF theory restricts star formation to the formation of embedded star clusters. Stars form  adhering to a universal form of the IMF but with the upper mass limit of each cluster set by its mass, while the clusters form with a consistent cluster mass function whose maximum mass is set empirically by the SFR of the galaxy \cp{weidner05,weidner06}. According to this theory, the galaxy-wide IMF is top-light for galaxies with a small SFR and top heavy at large SFR \cp{yan17}. Thus, the low stellar and gas densities found in the outer discs of galaxies provide an interesting extreme environment  for probing the nature of the IMF, and how it may vary with star formation intensity.

The correlation between gas density and SFR density was first proposed by \ct{schmidt59} as a power law in volumetric gas density $\rho_{\text{SFR}} \propto \rho_{\text{gas}}^n$ and predicted theoretically that $n = 1.5$. It is  commonly expressed in terms of the projected quantities  $\Sigma_{\text{SFR}} = A \Sigma_{\text{gas}} ^N$, which are more easily measured. \ct{kennicutt98} studied this correlation over a range of star-forming conditions with \HA\ derived SFRs and \HI\ + molecular (\HH) gas estimates, determining  $N=1.4 \pm 0.15$ and $A = (2.5 \pm 0.7) \times 10^{-4} M_{\sun}\  \text{yr}^{-1}\  \text{kpc}^{-2}$. This became the canonical "Star Formation Law" (SFL). However, recent determinations on both local and global scales have produced both steeper and flatter values of $N$  \cp[e.g.][]{momose12,roychowdhury15}.  Almost linear values of $N$ have been determined in the bright centres of galaxies where SFR is more strongly correlated with \HH \ surface density \cp[e.g.][]{bigiel08,leroy08,leroy13,pflamm08,pflamm092},  but in the \HI\ dominated,  extended outer discs of spirals and low mass dwarf galaxies SFR correlates strongly with the \HI\ surface density and $N$ is characteristically steeper, while molecular gas is undetectable  \cp[e.g.][]{bigiel08,bigiel10,wyder09, dz14, wang17}.  There are some empirically proposed complications to the SFL: such as truncated star formation due to disc stability \cp{kennicutt89, martin01}, and  a SFL that depends on both the stellar and ISM mass densities \cp{ryder94, dopita94,wang17}.

The theoretical models of \ct{ostriker10} and \ct{krumholz13} suggest  the scatter observed in determinations of $N$ in \HI\ dominated regions could be explained  by a  \HH\  gas density that is sensitive to other parameters such as metallicity and stellar and dark matter density. Determining the SFL in the outer discs of galaxies gives us a unique way of comparing these models to observations.

\begin{figure}
\includegraphics[width=0.5\textwidth]{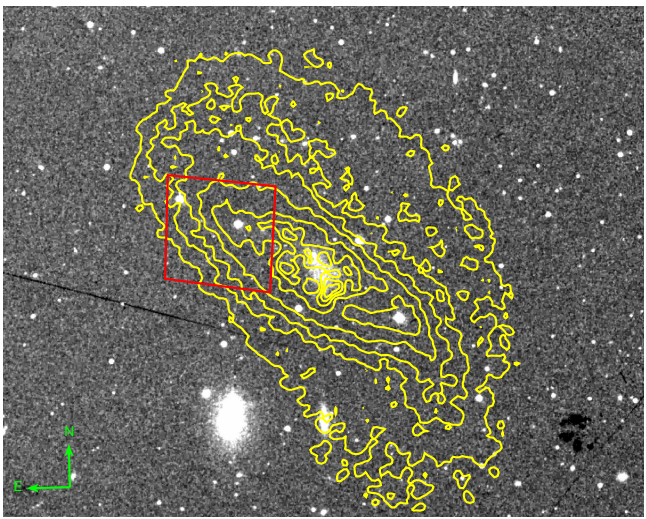}
\caption{ Palomar Observatory Sky Survey 103aE (red) digitised photograph of DDO 154 showing the central low surface brightness optical component of the galaxy with the \emph{HST}/ACS footprint overlaid in red. The image covers an area of 22$'$ $\times$ 16.5$'$.  Contours of \HI\ mass surface density from THINGS \cp{walter08} are overlaid in yellow at densities of 0.1, 0.5, 1, 2, 4, 6, 8, and 10 M$_{\sun}$ pc$^{-2}$.   \label{fig:composite}}
\end{figure}

In this work we constrain the IMF and SFL in the extended \HI\ outer disc of the local dwarf irregular galaxy DDO 154, using data from the Hubble Space Telescope and the Very Large Array.  DDO154 is one of the most gas rich galaxies known, harbouring a \HI\ disc that extends to $\sim 6$ times the Holmberg radius\footnote{DDO154 is projected on the sky  5.1 arcmin NW of the background Coma cluster galaxy NGC4789 (see Fig. \ref{fig:composite}), hence it is also catalogued as NGC489A.} \cp{kenn01} . Despite being so gas rich, its SFR is typical of those observed for dwarf irregular galaxies implying the star formation efficiency (SFE) is low \cp{kenn01}. Mass modelling of DDO 154 reveals a mass to light ratio  ($M_T / L_B$)  of 80, corresponding to a 90\% dark matter content, making it also one of the `darkest' disc galaxies known \cp{carig88, carig98,deblok08}. Since its nearest known neighbours are $>350$ kpc away \cp{carig98}  tidal interactions (i.e. external triggers) are not a likely cause for these extreme properties nor for any postulated  recent fluctuations in its star formation rate. This makes it a valuable candidate for investigating  the SFL and variations in the IMF, and Table \ref{tab:gal_params} lists the adopted parameters for DDO 154 and their source. We adopt the upper mass limit of $M_U = 120 M_{\sun}$ and high mass slope $\alpha = -2.35$ as the standard IMF and refer it to as the Kroupa IMF \cp{kroupa01}.   This paper  is organised as follows. In  \S\ref{sec:2} we present our data and photometry. In \S\ref{sec:3} we describe our modelling of the CMD and  statistical testing of it, and in \S\ref{sec:4} we present out results constraining the IMF. In \S\ref{sec:5} we determine the SFL in DDO154 and  investigate the stability of the gas disc, and in \S\ref{sec:6} we conclude.

\begin{table}
 \caption{Adopted properties for DDO 154 \label{tab:gal_params}}
\resizebox{0.5\textwidth}{!}{%
 \begin{tabular}{| l | c | l | }
 \hline
Parameter &   Value & Reference  \\ \hline \hline
Metallicity (Z) & 0.1 Z$_{\sun}$ & \ct{kenn01} \\ 
Distance (Mpc) & 4.02   & \ct{jacobs09} \\ 
Environment & Isolated & \ct{karachentsev04} \\
Pixel Scale (pc/pix) & 0.974 & - \\ 
$E(B-V)^G$ (mag) & 0.009 & \ct{jacobs09}  \\ 
Inclination & $66^\circ$  & \ct{walter08} \\ 
Position Angle & $229.7^{\circ}$ & \ct{deblok08}\\
V$_{\text{hel}}$ (km s$^{-1}$) & 374 & \ct{karachentsev13} \\ 
M$_{\text{\HI}}$ ($10^7 \, M_{\sun}$) & 39.4 & \ct{carig98} \\ 
M$_B$ (Mag) &  -13.92 & \ct{karachentsev13}   \\ 
$M_T / L_B$ & 80 & \ct{kenn01}   \\ 
$M_{\star} (10^7 M_{\sun})$ & 1.26 & \ct{leroy08} \\
SFR  (\HA, $10^{-3}  \, M_{\sun}$ yr$^{-1}$) & $ 2.79 \pm 0.56$ & \ct{kenn01}   \\
SFR (UV + IR,  $10^{-3}  \, M_{\sun}$ yr$^{-1}$) &$ 5 $& \ct{leroy08}\\
\hline
 \end{tabular}}
\end{table}

\section{Data and Analysis} \label{sec:2}
\subsection{\emph{HST} images}
We use data from the Advanced Camera for Surveys (ACS) Wide Field Camera (WFC)  of the Hubble Space Telescope \emph{HST} (proposal ID:10287). The observations consist of three exposures in the F475W ($g_{475}$), F606W ($V_{606}$), and F814W ($I_{814}$) filters with combined exposure times of 1800, 1800, and 3433  seconds respectively. The observations focus on the extended outer disc centred on  R.A.: 12$^h$ 54$^m$ 19.9$^s$, Dec:+27$^d$  10.2$^m$ 11.2$^s$ (J2000), shown in Fig. \ref{fig:composite}, 5.71 kpc from the centre of the galaxy and  covering radii between $98 - 340$ arcsec ($2.3-12.2$ kpc, corrected for inclination and position angle).

\subsection{Photometry}
We perform PSF photometry  using  \dolphot, a modified version of HSTphot designed specifically  for ACS/WFC images \cp{dolphot}.  Preprocessing is done using the \dolphot\ routines \emph{acsmask}, \emph{splitgroups} and \emph{calcsky} to apply the pixel masks, separate the two  ACS  chips, and produce robust sky images respectively. We follow the method of \cta{sarah15} and adopt the recommended parameter file from the \dolphot\ ACS user manual with changes to a few select parameters as suggested by  \ct{dalcanton09}. We set the photometry parameters \emph{FitSky}, \emph{RAper} and \emph{Force1} to their recommended values as they were found to have the greatest effects on the quality of the photometry.  \dolphot\ outputs all detected objects with positions on the reference image and magnitudes in the VEGAMAG system calculated using the transforms of \ct{sirianni05} and corrected for CTE loss  following \ct{riess04}. We use the photometric quality parameters roundness, sharpness, crowding, and signal to noise (S/N) provided for each star in each filter  to clean the catalogue of non-stellar detections, adopting the same  quality cuts  as \cta{sarah15}  and \ct{dalcanton09}. To make the final photometric catalogue stars must satisfy (S/N)$_{1,2} \ge 4$, |sharp$_{1}$ + sharp$_{2}$| < 0.05, (crowd$_{1}$ + crowd$_{2}$) < 0.6, and  |round$_{1}$ + round$_{2}$| < 1.4 where the subscripts $(1,2)$ denote one of the filter pairs $(g_{475}, V_{606})$ or $(V_{606}, I_{814})$. Satisfying the conditions in at least one filter pair was sufficient to be included in the final photometric catalogue. The final catalogue consisted of 2698 stars out of the  173 112 in the original  \dolphot\ output.

\begin{figure}
\includegraphics[width=0.45\textwidth]{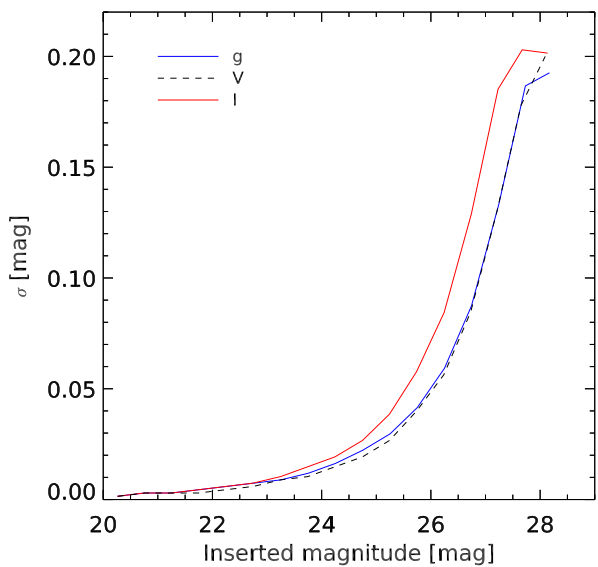}
\caption{Median photometric uncertainties derived from artificial star tests. \label{fig:mad}}
\end{figure}

\begin{figure*}
\includegraphics[width=0.45\textwidth]{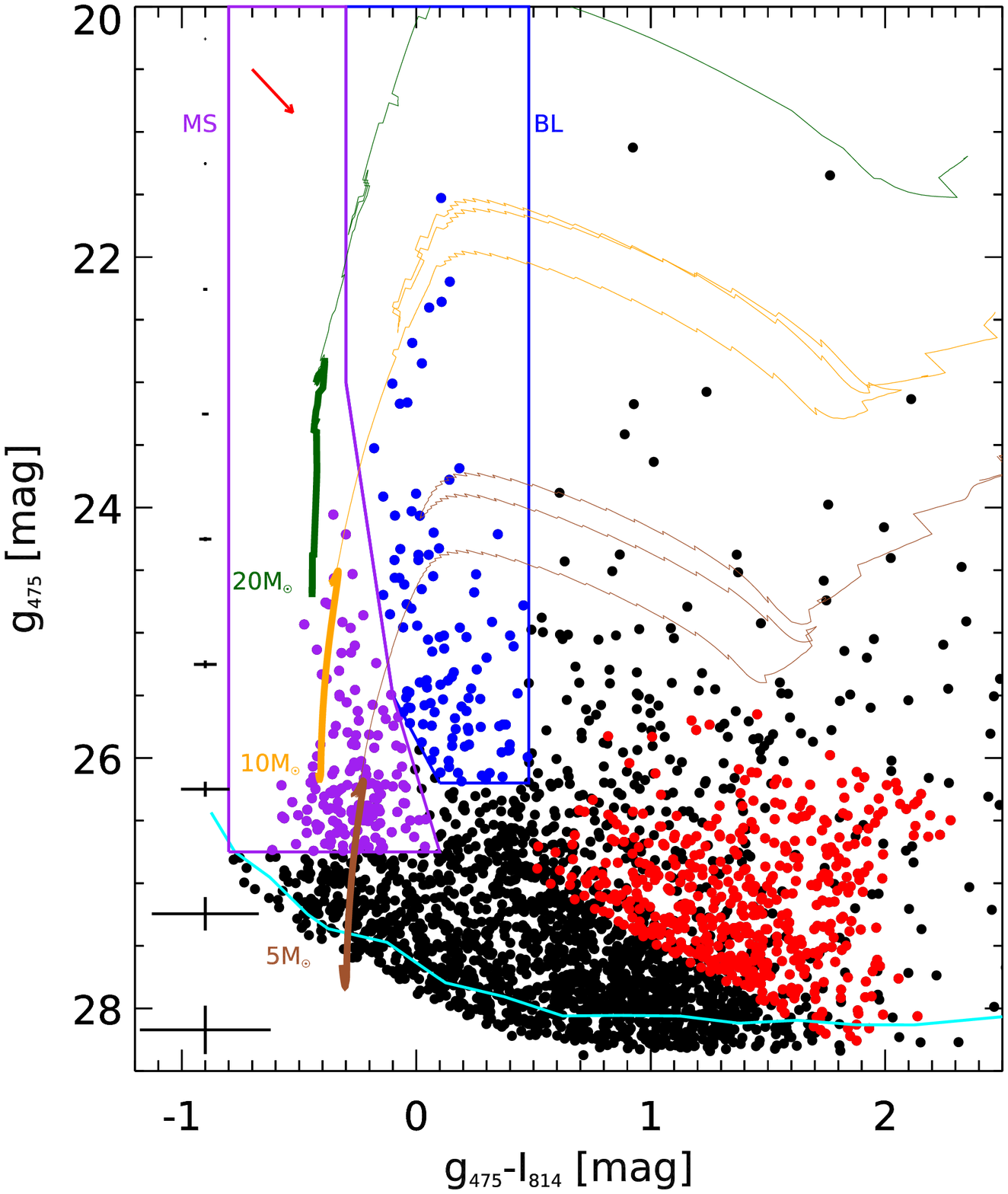}
\includegraphics[width=0.45\textwidth]{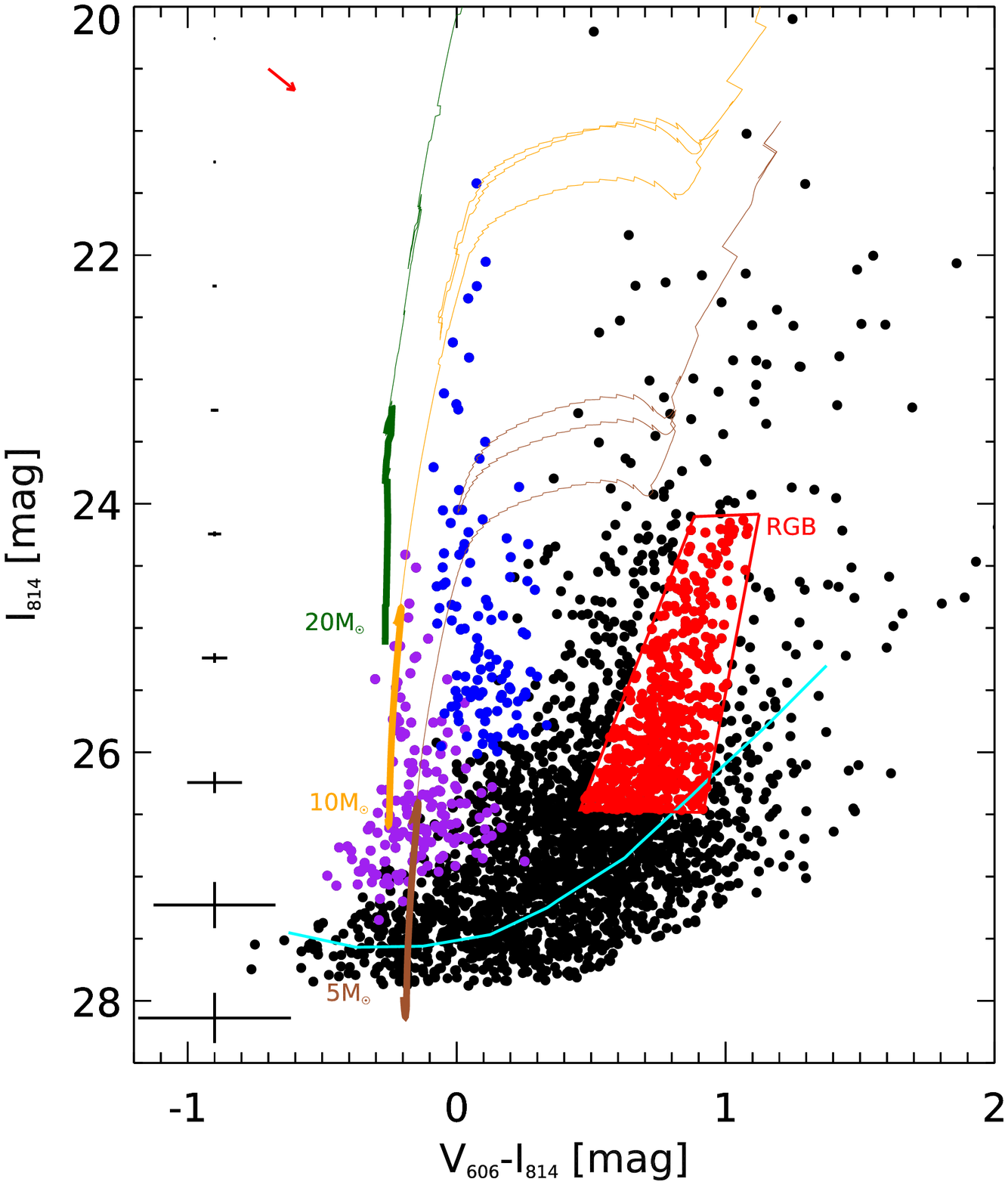}
\caption{CMDs of the stars recovered from the \emph{HST} ACS/WFC images in $g_{475}$ vs. $(g_{475} - I_{814})$ (left)  and  $I_{814}$ vs. $(V_{606} - I_{814})$ (right). Error bars on the left show photometric uncertainties and the cyan lines show the 60\% completeness limit derived from artificial star tests.  PARSEC evolutionary tracks \cp{bressan12,rosenfield16} for 5, 10, and 20 $M_{\sun}$ evolutionary tracks are shown in brown, orange and dark green respectively with thicker sections corresponding to the MS phase. The red arrow in the top left of each CMD  corresponds to the reddening $E(B-V) = 0.1$, ten times the adopted value for DDO 154. The $g_{475} - I_{814}$ (left) CMD shows the MS and BL selection boxes  in purple and blue respectively. The $V_{606} - I_{814}$ CMD (right) outlines the RGB selection box in red. Stars in the MS, BL and RGB selection boxes are coloured purple, blue, and red respectively in both panels. \label{fig:cmd}}
\end{figure*}
\subsection{Artificial star tests} \label{subsec:ast}
To quantify photometric uncertainties and the completeness of the CMDs we ran artificial star tests using \dolphot\ (similar to the work of \citealt{dalcanton12} and  \cta{sarah15}). The inserted and recovered magnitudes of these stars determines how well objects of known magnitude and colour are recovered by \dolphot. We simulated 300,000 artificial stars with colour and magnitude limits set by the observed data, distributed uniformly over the images as crowding is low in the outer disc. The colour limits were defined using  the $g_{475}$ and $I_{814}$  bands  since they provide the largest colour baseline and we account for scatter due to photometric uncertainties ($-1 \le g_{475} - I_{184} \le 3$ and   $20 \le g_{475} \le 29$). This produces a catalogue of the artificial stars with their inserted and recovered magnitudes, positions, and quality parameters to which we applied the same quality cuts. Using Gaussian statistics the median absolute deviation between the inserted and recovered magnitude of stars in 0.5 mag bins is calculated to quantify the photometric uncertainties, which we show in  Fig. \ref{fig:mad}  as a function of inserted magnitude. The completeness of the CMD is quantified by comparing the ratio of the number of surviving  to inserted artificial stars as a function of position of the CMD using bins of 0.25 mag and 0.5 mag in colour and magnitude respectively. The cyan completeness contour in Fig. \ref{fig:cmd} marks the region of the CMD below which less than 60\% of inserted artificial stars are recovered in the same bin. The features of the CMD we are interested in are above this contour, and so only mildly affected by incompleteness. We use the photometric uncertainties and completeness map  to correct our  stellar population models and account for these effects.

\subsection{Final photometric catalogues}
The CMDs  shown in  Fig. \ref{fig:cmd} reveal the different populations present in the outer disc of DDO 154. Aside from the red giant branch stars, which formed between 1 and 10 Gyr ago, the most prominent feature is the vertical sequence of stars rising around colour $(g_{475} - I_{814}) = 0$. This is a mixture of young main sequence (MS) stars  and older, lower mass helium burning `blue loop' (BL) stars  \cp{dalcanton12}. As these stars could contaminate our  MS selection, we take careful considerations to avoid them. The largest colour separation is provided by the $(g_{475} - I_{814})$ colour CMD, better disentangling the bluer MS stars from the redder BL stars. This is supported by the evolutionary tracks shown in Fig. \ref{fig:cmd}: the thicker part of the tracks (MS phase) is consistent with the bluer portion of the vertical sequence, while the redder half is consistent with vertex of the tracks migrating left after the RGB  phase. Considering the photometric uncertainties we separate the two populations using the selection boxes shown in Fig. \ref{fig:cmd} to minimise  contamination from BL stars. There are 162 MS stars in our selection box,  the faintest of which roughly corresponds to the MS turn-off luminosity of a  $\sim 4 \ M_{\sun}$ star.

\begin{figure*}
\includegraphics[width=0.45\textwidth]{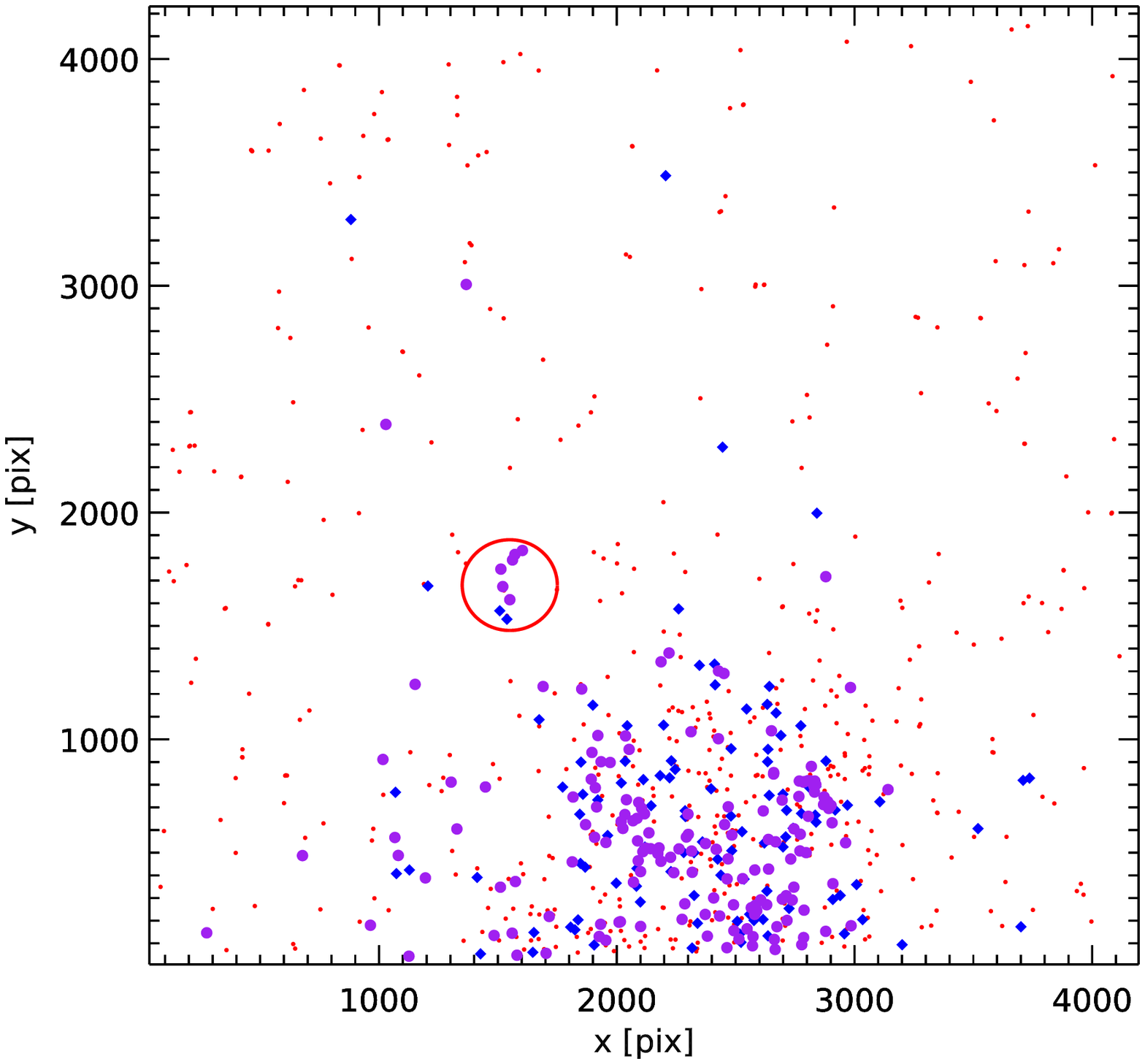}
\includegraphics[width=0.45\textwidth]{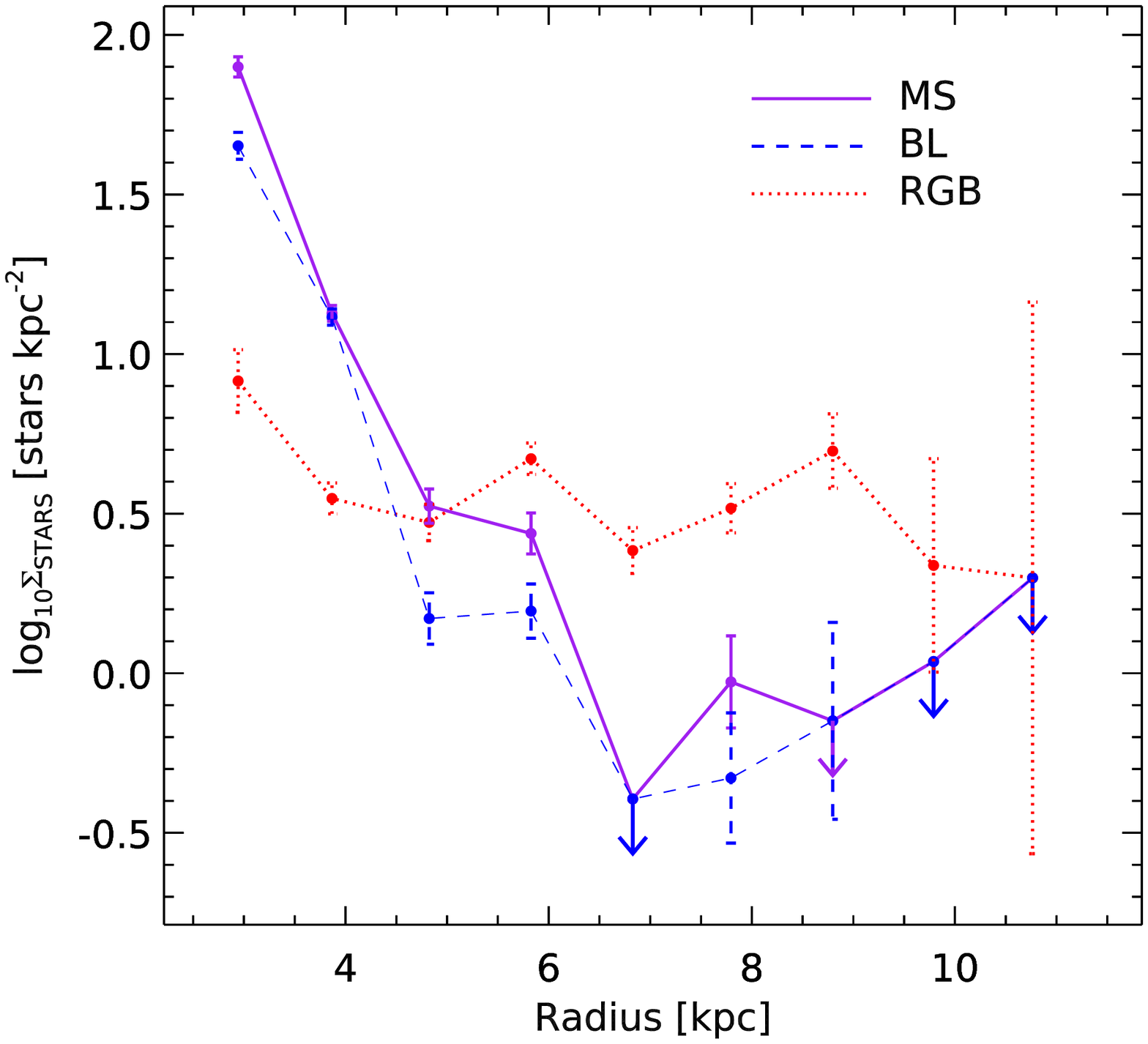}
\caption{(Left) Spatial distribution of recovered stars in the \emph{HST} images. RGB stars are shown in red, MS stars in purple, and BL stars as blue diamonds. (Right) Disk-plane radial distribution of MS (purple), BL (blue, dashed) and RGB (red, dotted) stars. Error bars correspond to the uncertainty calculated from Poisson  statistics, and downward arrows denote upper limits of 1/Area for empty radial bins.   \label{fig:spatial}}
\end{figure*}

In the right panel of Fig. \ref{fig:spatial} we plot the spatial distribution of the stars, colour coded by their evolutionary phase. The RGB stars, aside from some clustering near the main stellar disc at the bottom of the image,  are distributed relatively uniformly over the field.  The MS and BL stars have a similar but noticeably more clumpy distribution.  This is shown in the right panel of Fig. \ref{fig:spatial}  where we plot the disk-plane radial distribution of stars in bins of 0.5 kpc. The surface density of RGB stars shows a gradual decline with radius, whereas the MS and BL stars show a sharper decline. Aside from one small association not far from the main stellar component, there are almost no MS or BL stars in the extended outer disc, implying recent star formation is almost absent. This association (hereafter ASN-1) is marked in the left panel of Fig. \ref{fig:spatial} with a red circle and is located at R.A.: 12$^h$ 54$^m$ 18.4$^s$, Dec:+27$^d$  9$^m$ 41.3$^s$ (J2000),  which is 5.58 kpc from the centre. It has a maximum radius of $\sim195$ pc,  and contains 6 MS stars and 2 BL stars.

\subsection{Comparisons to GALEX UV images}

\begin{figure*}
\includegraphics[width=0.45\textwidth]{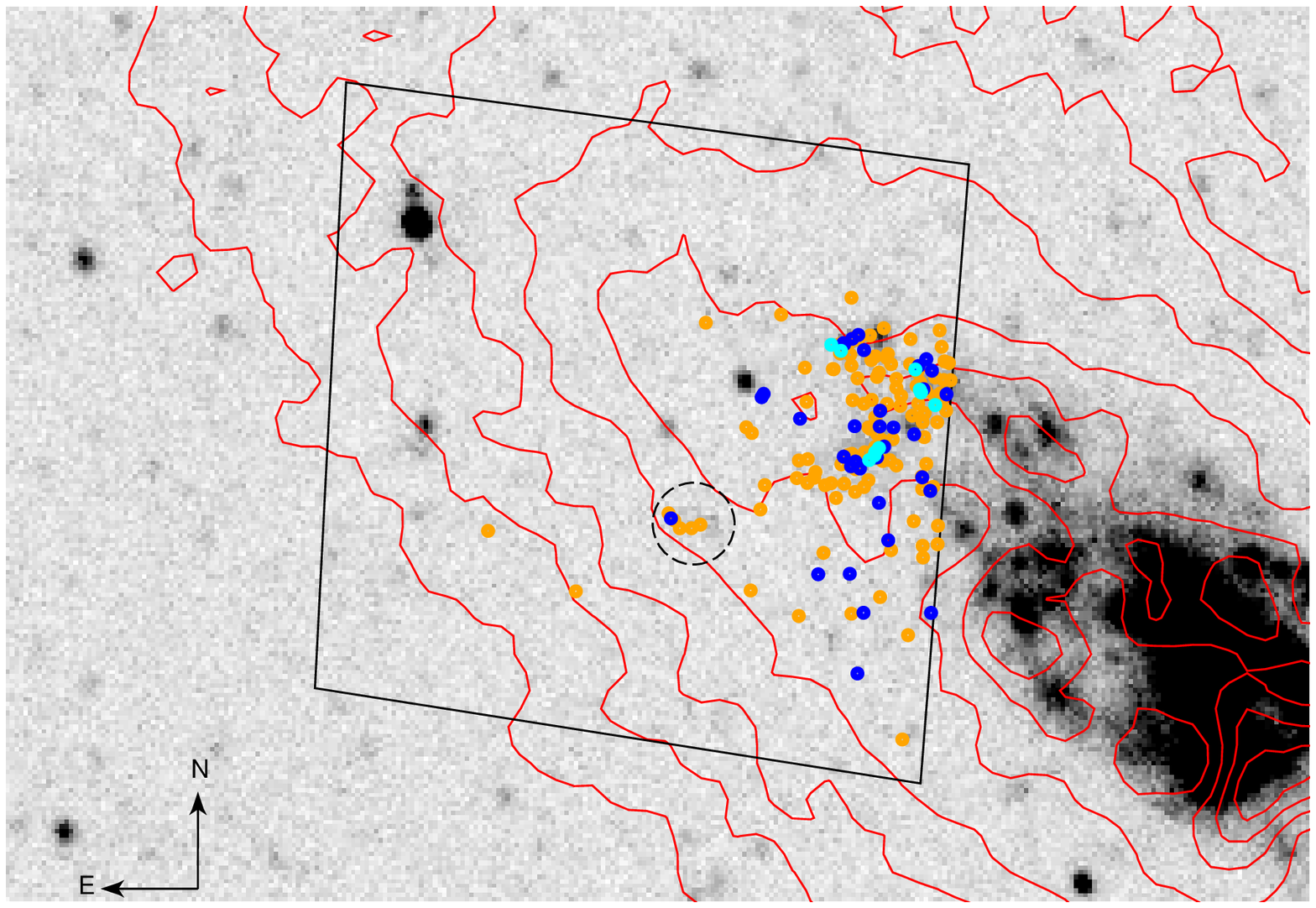}
\includegraphics[width=0.45\textwidth]{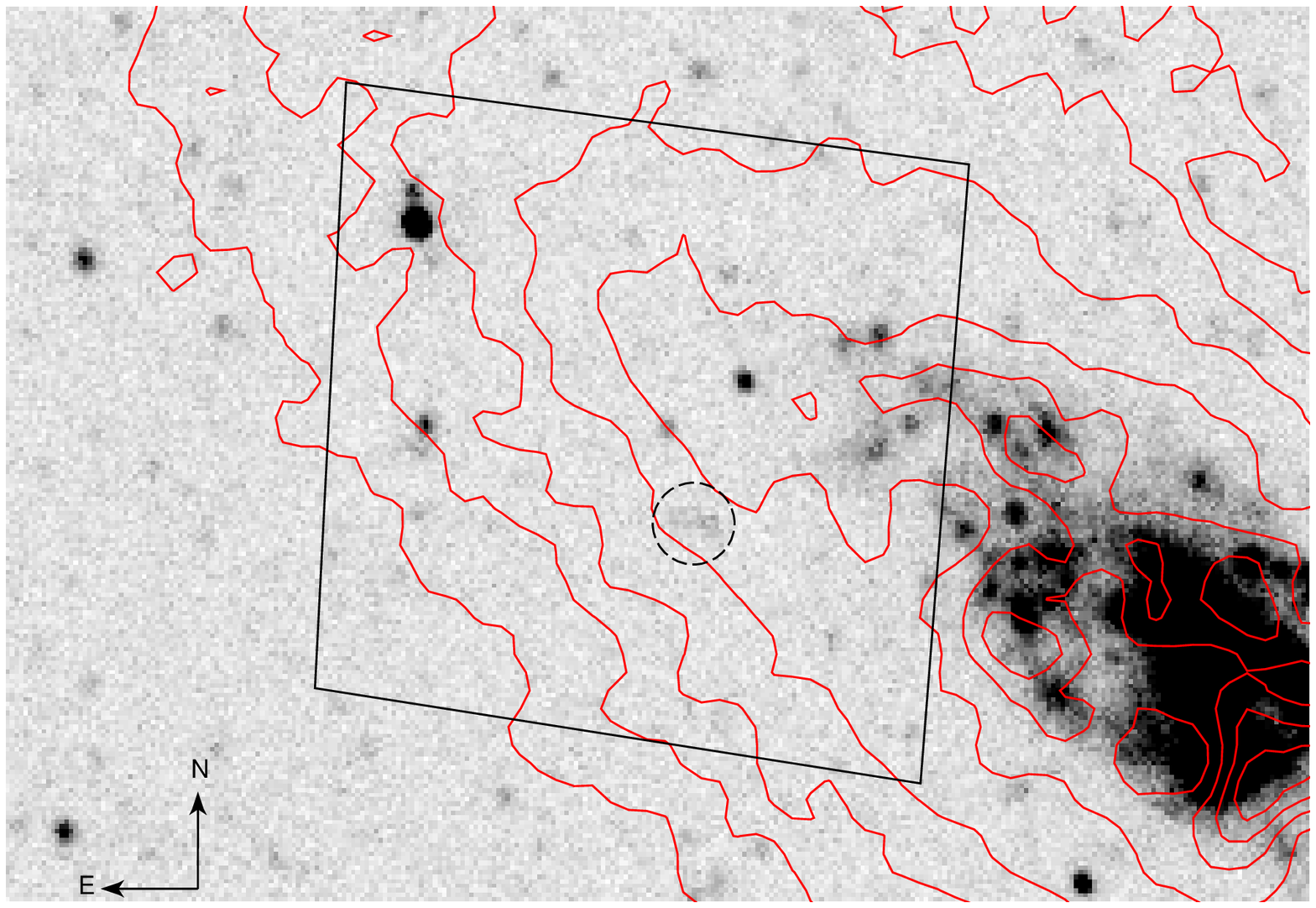}
\caption{Comparison between the GALEX NUV image   and the location of MS stars detected in the \emph{HST} images of the outer disc of DDO 154. Stars with $g_{475} \geq 26$ mag are shown in orange, $26 < g_{475} \leq 25$ mag in blue, and $g_{475}<25$ mag in cyan.  We include images with (left) and without (right) MS detections to reduce crowding and make the  NUV emission  clearer.  The \emph{HST} footprint is shown in black, and the image orientation is shown by the compass in the bottom right corner.  Some individual stars appear to show no correlated detection in GALEX, and faint emission can be seen associated with ASN-1 marked by the dashed circle. The brightest NUV sources in the field of view with no correlated MS star detection are foreground stars.  \label{fig:uv}}
\end{figure*}

As the stars in our MS selection box are more massive than $4 M_{\sun}$ (i.e. B and O stars) we  should observe emission in the UV detectable by the GALEX satellite. In Fig. \ref{fig:uv} we  compare the locations of the  MS stars to  NUV images of DDO 154. The majority of the UV emission is located near the main stellar disc. There is good correlation between the MS stars and UV emission, but some stars are too faint or sparsely distributed to be detected by GALEX. Of the MS stars at $R>5$ kpc only ASN-1 shows correlated UV emission.

\section{Stellar population models} \label{sec:stelpop} \label{sec:3}
\subsection{Star formation history} \label{subsec:sfh}
Either a recent drop in the SFR (a "gasp" in star formation) or a deficiency of high mass stars in the IMF may cause a deficiency of bright blue stars. Thus the IMF and SFH are degenerate; we cannot use our data to solve for both  \cp{miller79}. Here we select a sensible form for the SFH  to break this degeneracy and infer the  IMF from our data. Star formation in dwarf irregular galaxies is commonly assumed to be continuous and low level, with intermittent bursts \cp{gerola80} triggered by gaseous inflow or interactions \cp{taylor93,taylor95}. Starbursts are commonly assumed to be short lived, with timescales as short as a few Myr \cp{ferguson98,tremonti01,harris04_2}, but recent observations   have shown that star formation can be significantly elevated on whole-galaxy scales over $\sim100$ Myr to Gyr timescales \cp{mcquinn10a, mcquinn10b, mcquinn12}.  This agrees with crossing time arguments where the duration of star formation should not be shorter than the time it would take a disturbance (such as a supernova explosion) to cross the star-forming region and quench further star formation \cp{meurer00}. This essentially restricts the rapid quenching of star formation to local scales while the global SFR of the galaxy remains elevated and constant. Star formation rates elevated on Gyr timescales have been produced in controlled, non-cosmological  simulations of  low mass galaxies \cp[e.g.][]{verbeke14, watts16}.  Cosmological zoom-in simulations have produced dwarf galaxy SFHs consisting of  consecutive bursts that are effectively constant over Gyr timescales \cp[e.g.][]{onorbe15}, and centrally concentrated $\sim5$ to $\sim50$ Myr duration bursty SFHs with longer periods of quiescence \cp[e.g.][]{sparre17}. 

DDO 154  is an isolated galaxy \cp{karachentsev04, read17}, its nearest neighbours are NGC 4826 and UGC 7698, both with projected distances greater than 350 kpc away  \cp{carig98} indicating there has been no recent external triggers capable of creating a starburst. It is undergoing low intensity star formation, harbouring several small \HII\ regions in its main stellar component. The combined \HA\ equivalent width from spectrophotometry is  $31 \pm 6 \ \AA$ \cp{kenn01} and most of its stellar mass is thought to have been built up by this low intensity star formation. At the current UV+IR SFR of $5\times 10^{-3} \  M_{\sun} \ \text{yr}^{-1}$ \cp{leroy08} it would take 2.52 Gyr to build the estimated stellar mass of DDO 154 \cp[$1.26 \times 10^{7} \ M_{\sun}$,][]{leroy08}.   We are not concerned with the central parts of the galaxy, but instead the extended outer disc where DDO 154 shows increasing disk stability, especially beyond $R = 5$ kpc \cp{meurer13}.  This may mean it is harder for gas clouds to collapse and stars to form.

The MS stars in our images  span the majority of the field of view so the minimum timescale we would expect star formation to have lasted would be the crossing time of the ACS/WFC images. Assuming stars are travelling at the typical \HI\ velocity dispersion of 8.62 km s$^{-1}$ (the mean value interior to the radius of 7.5 kpc in our \emph{HST} field of view), this corresponds to an image crossing time of 462 Myr. The lowest mass MS star in our selection box is 4 $M_{\sun}$ which has a MS lifetime of 160 Myr \cp{schaerer93} making this the longest timescale we are sensitive to. As the crossing timescale estimate is over twice  this MS lifetime, a constant SFR is a reasonable assumption. We also consider the possibility of a declining SFR due to the consumption of gas based on the model of \ct{kennicutt98}, where the SFR decreases by 10\% each dynamical timescale. The centre of our \emph{HST}/ACS images is 5.71 kpc from the centre of the galaxy, which using the THINGS rotation curve \cp{deblok08} corresponds to $v_{circ} = 47.2$ km s$^{-1}$.  Assuming circular orbits, the orbital time is 743 Myr and dynamical time  242 Myr, corresponding to a 6.7\% decrease in SFR over the lifetime of a 4 M$_{\sun}$ star. \cta{sarah15} investigated declining star formation rates of similar magnitude and timescale, and found them to produce results effectively equivalent to a constant SFR.

In addition to temporal variations in the formation of stars; their formation environment could also play a role in variations in the high mass end of the IMF. There are two competing models for the formation of high mass stars: competitive accretion and monolithic collapse. In the competitive accretion model  \cp[e.g.][]{bonnell02} high mass stars are formed through interactions between protostars and the accretion of gas in  dense environments. This restricts their formation to star clusters where densities are high. The collapse model \cp[e.g.][]{yorke02} determines the formation of high mass stars from the properties of the parent molecular cloud and  allows high mass stars to form anywhere. Following the review of \ct{lada03} it is commonly assumed that star formation  is restricted to clusters,  however this assumption may not be valid in low density environments.  The review of \ct{krumholz14}  notes that the fraction of stars formed in clusters depends strongly on the density thresholds used to define them \cp[cf. ][]{bressert10}.  The broad range of densities in which protostars are found neither excludes nor rules out star formation occurring only in clusters \cp{gieles12}.  However,  the models of \ct{kruijssen12} indicate that low density star-forming environments should not contribute  significantly to bound star formation.   \ct{gutermuth11}   showed that the  formation of protostars over large ranges in gas column density shows no distinct break between cluster and field population \cp[cf.][]{megeath16}. While \ct{stephens17} find that there are no truly isolated O stars in the Magellanic Clouds, and \ct{wright18} show OB associations with substructures,  \ct{elmegreen08} suggested that O and B stars form in loose associations rather than bound clusters in low gas pressure environments. At higher star formation intensities,  \ct{meurer95} and \ct{larsen00} have shown that $\sim20$\% of the UV light from starburst galaxies comes from young compact star clusters, with a smaller contribution at lower star formation intensity, consistent with other observations \cp{bressert10}  and models \cp{kruijssen12} that find $\lesssim 35 \%$ of stars form in dense bound clusters. This implies that the formation of bound star clusters should not be prevalent in low density regions such as the outer disc of DDO 154. 

As the simplest approach we  modelled  the outer disc of DDO 154 as random sampling of the IMF and a constant SFH  over the dynamical timescale without clustering in space or time. 

\subsection{Simulated CMDs}
To constrain the IMF we compare the main sequence luminosity function (MSLF) of the stellar population in DDO 154 to our simulations.  Model CMDs are created by randomly distributing 500,000 stars over an adopted IMF with a constant SFR over 400 Myr. The simulation time  is selected to be longer than the MS lifetime of the lowest mass star in our selection box (160 Myr) to account for  lower mass stars that could be scattered into the selection box by photometric uncertainties. A lower mass limit of 2.5 $M_{\sun}$ is selected to ensure the simulated CMD matches observations as closely as possible and  to account for the photometric scattering of fainter stars into the selection box. We adopt the common assumptions for modelling a resolved stellar population: stars are non-rotating, have zero binary fraction, and a uniform metallicity and dust extinction \cp[e.g.][]{annibali13,hillis16,sacci16}.  We explore the limitations of these assumptions in  \S\ref{subsec:caveats}.

The metallicity $Z=0.0013$ is estimated  from the Oxygen abundance $12+\log(O/H) = 7.67 \pm 0.05$ \cp{kenn01} using the \ct{asplund09} abundance scale, and is approximately 10\% of the Solar value. We use the PARSEC interpolated evolutionary tracks \cp{bressan12, rosenfield16} with metallicity $Z=0.001$ (closest to our adopted value) covering the mass range $1 < m/M_{\sun} < 120$ and evolution of stellar properties from the pre-MS to the end of the asymptotic giant branch. We interpolate between evolutionary tracks for the effective temperature ($T_{\text{eff}}$) and surface gravity ($\log g$) of each star based on its initial mass and age at the end of the simulation. Synthetic photometry is created from the available grid of ATLAS9 spectra \cp{castelli04} using {\sc pysynphot}, and interpolated to the $T_{\text{eff}}$ and $\log g$ values of the simulated stars.  Stars in the pre-MS phase or those too old to be covered by the evolutionary tracks are flagged and removed as they do not meaningfully contribute to the areas of the CMD we are interested in.  The model photometry was corrected to the distance and foreground Galactic dust content (Table \ref{tab:gal_params}). Photometric uncertainties are modelled by perturbing magnitudes by a Gaussian random variate with a mean set by the star's magnitude and  standard deviation set by the median absolute deviation calculated from artificial star tests (\S\ref{subsec:ast}, Fig. \ref{fig:mad}). Completeness of the CMD is modelled by employing a survival analysis  where stars are randomly removed using the fractional completeness map produced from the artificial star tests (\S\ref{subsec:ast}). The simulated populations were created in a grid of two parameters; the IMF upper mass limits $M_U / M_{\sun} = 10, 16, 20, 30, 40, 50, 60, 70, 80, 90, 100, 120$ and high mass slopes $-3.95 \leq \alpha \leq -1.95$ in steps of 0.1.  An ensemble of 100 simulated populations are created for each pair of IMF parameters $(M_U,\alpha)$ to measure stochastic effects, and determine more robust mean results. 

\section{MS luminosity function constraints on the IMF} \label{sec:4}
\subsection{Best-fitting IMF}

We use the two sided Kolmogorov-Smirnov (KS) test to compare the cumulative distribution function (CDF) of the $g_{475}$ band MSLF of the outer disc of  DDO 154 to the CDF of stars randomly selected from our simulated stellar populations, as previously done by  \cta{sarah15}. The number of simulated stars in our MS selection box varies, depending on the IMF parameters, between $\sim 1400$ ($\alpha$ = -3.95, $M_U$ = 10 $M_{\sun}$) to $\sim 4300$ ($\alpha$ = -1.95, $M_U$ = 16 $M_{\sun}$).   We randomly select $N$ simulated stars from each simulation for comparison where $N$ is a random variate drawn from a Poisson distribution with a mean equal to the number of observed MS stars (162). The KS test parameter  $D$, which represents the maximum deviation between the CDFs of the observations and simulations, is averaged over the 100 simulations for each point in the $(M_U,\alpha)$ parameter space. The simulation set with the minimum $D_{\rm ave}$  is then taken to be the best-fitting pair of IMF parameters.

\begin{figure}
\includegraphics[width=0.45\textwidth]{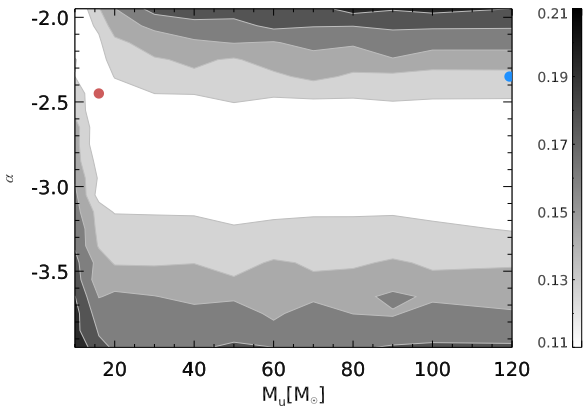}
\caption{Contour plot of mean KS test $D$ statistic across the $(M_U,\alpha)$ parameter space. The red point corresponds to the best-fitting IMF parameters, and the blue point to a Kroupa IMF. \label{fig:contour}}
\end{figure}

\begin{figure}
\includegraphics[width=0.45\textwidth]{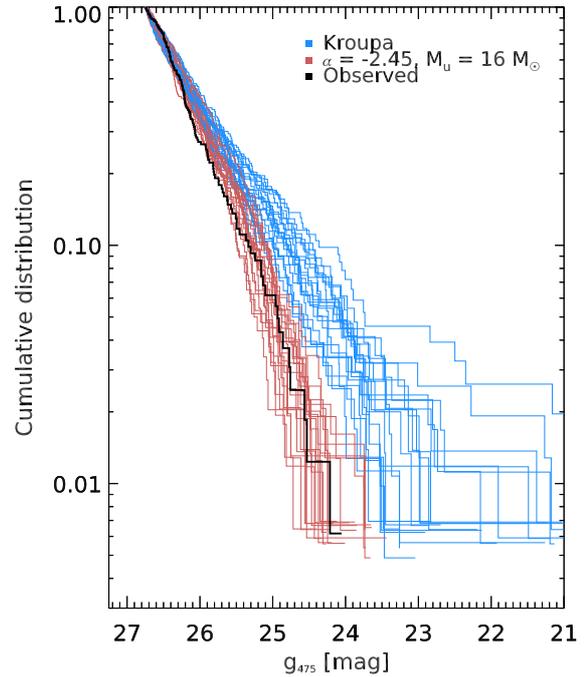}
\caption{Comparison of cumulative MSLFs between that observed in DDO 154 (black) and the 20 best-fitting simulations from a Kroupa IMF (blue) and best-fitting IMF (red). \label{fig:compare}}
\end{figure}

Fig. \ref{fig:contour} shows the  distribution over the parameter space  $D_{\rm ave}$ values  as a contour plot. The best-fitting IMF parameters are $\alpha$ = -2.45 and $M_U$ = 16 $M_{\sun}$  which are referred to as  the best-fitting IMF from here. There is an extended region of local minima spanning $16 < M_U / M_{\sun} < 120$  implying the upper mass limit is not well constrained.  Fig. \ref{fig:compare} compares the cumulative MSLF of the observed population (black) to the 20 best-fitting simulated populations from both a Kroupa IMF (blue) and best-fitting IMF (red).  Each simulated MSLF was produced using $N$ randomly selected stars, where $N$ is drawn from a Poisson distribution with mean of 162. The best-fit parameters clearly match  the observed data  better than a Kroupa IMF for a stellar population with a constant SFH  over the dynamical timescale. If the Kroupa IMF was a good prescription for star formation in the outer disc of DDO 154, we would expect $\sim$10\% ($\sim$1\%) of stars to be brighter than $g_{475} \sim 24.8$ (23.3) mag. Instead, the observed population shows $\sim$10\% ($\sim$1\%) of stars brighter than $g_{475} \sim$ 25.3 (24.2) mag. Hence if stars were forming in the outer disk with a Kroupa IMF, we would see relatively more at brighter magnitudes. Table \ref{tab:d_p} shows the  KS test results   between  the observed MSLF and  the Kroupa and best-fitting IMF  averaged  over the 100 simulated stellar  populations, and the results of the single best-fitting simulation from each IMF. The  $D_{\rm ave}$ value  disfavours the Kroupa IMF which yields a $p=0.088$ significance level. 

 In \S\ref{subsec:sfh} we provided justification for our assumption of a constant SFH, but have not provided any sanity checks to confirm that this assumption is correct. The population of BL stars can be used to provide such a check. Our first sanity check compares the cumulative luminosity function of observed BL stars to simulation populations. Like the MSLF analysis, the Kroupa IMF simulations over-predict the number of BL stars compared to the observations. The best-fit IMF simulations are a closer match the BL luminosity function, but slightly under-predict the number of bright BL stars. Our second sanity check compares the ratio of BL to MS stars to the observed fraction BL/MS = 0.660. The average BL/MS ratio of the 100 simulations assuming a best-fit and Kroupa IMF are $0.644\pm 0.017$ and $0.571 \pm0.015$ respectively, where the uncertainties are the standard deviations. Clearly, best-fit IMF better reproduces the observed MS and BL stars under the assumption of a constant SFH.

\begin{table}
 \caption{KS test  maximum deviation ($D$)   averaged over the 100 simulations of a best-fitting IMF (first row) and a Kroupa IMF (second row) with corresponding $p$-values. $D_{\text{min}}$  and $p_{\text{min}}$  are also given for the single best-fitting MSLF from each set of simulations.  \label{tab:d_p}}
 \begin{tabular}{| c | c | c | c | c | }
 \hline
$(M_U,\alpha)$   &  $D_{\rm ave}$ & $p_{\rm ave}$ & $D_{\text{min}}$ & $p_{\text{min}}$  \\ \hline \hline
$(16 \ M_{\sun} , -2.45)$   & 0.108 & 0.288  & 0.065 & 0.884\\ 
$(120 \ M_{\sun}, -2.35)$   & 0.136 & 0.088 & 0.096 & 0.438  \\
\hline
 \end{tabular}
  \end{table}
  

\subsection{Uncertainties in IMF parameters}
\begin{figure}
\includegraphics[width=0.45\textwidth]{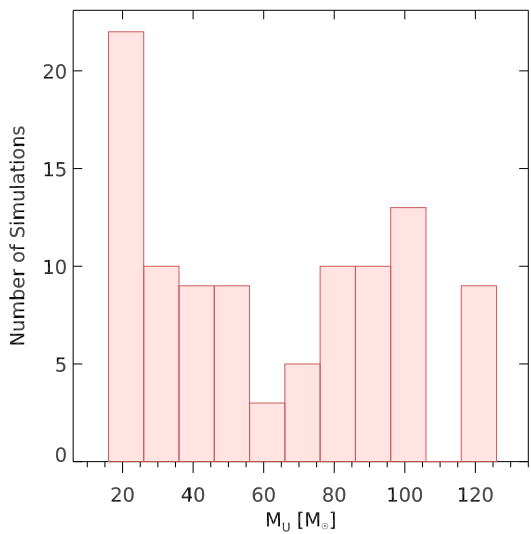}
\caption{Histogram of the 100 recovered $M_U$ values from KS tests of mock observations created from the best-fit IMF simulations. \label{fig:mu_hist}}
\end{figure}

We ran simulations to determine how well our method recovers  an IMF of  known input and the uncertainty on the derived parameters.  We generated mock observations by randomly selecting stars equal in number to the observed MS (162) from the  100 Kroupa and best-fitting IMF simulations. The KS test comparison is then rerun for each mock population against the entire grid of models to determine the best-fitting IMF parameters, and we use the mean and standard deviation of the  distribution of recovered parameters to quantify the uncertainties. For an input Kroupa IMF the average recovered parameters are $\alpha = -2.31 \pm0.29$ and $M_U = 61 \pm 35 \ M_{\sun}$, and for an input best-fitting IMF $\alpha = -2.58 \pm 0.37$ and $M_U = 36 \pm 29 \ M_{\sun}$. However, only the distribution of recovered $\alpha$ values for the best-fitting IMF are Gaussian; the uncertainty in $M_U$ is more lopsided as shown in Fig. \ref{fig:mu_hist}.  The large uncertainties quoted are consistent with the width and length of the region of local minima observed in Fig. \ref{fig:contour}.

\subsection{Caveats} \label{subsec:caveats}
Here we consider the assumptions we have made when modelling CMDs, and the effects they could have on our results.      

The evolutionary tracks we use do not include the effects of stellar rotation, which causes stars to have longer MS lifetimes, higher effective temperatures, rates of mass loss \cp{ekstrom12}, and by the MS turnoff, a higher luminosity \cp{potter12}. If our simulated stars turn off the MS early compared to the observed population, then there will appear to be an over-abundance of high mass stars, requiring IMF parameters that are flatter or have a higher upper mass limit to match the observations. This is also the case for the luminosity of the stars, if the non-rotating stars are fainter, then our simulations would require more high mass stars to match the observations, biasing us toward flatter IMF slopes and higher upper mass limits.  

We do not model the effects of binary stellar evolution \cp{eldridge17}, nor the presence of binary systems or star clusters. Typical binary  separations are $\sim 10$ AU \cp{minor13}  which are unresolved at the pixel scale 0.974 pc/pix of the images; effectively all binaries are unresolved in our data.  Similarly, there could be unresolved low mass clusters in the images appearing as bright stars.  \ct{hill06} find the average core radius of LMC clusters to be 4.2 pc which would be resolved by our images,  but some of their more compact observed clusters have core radii of just 0.61 pc which would be unresolved. \ct{weidner09} investigated the effects of unresolved stellar multiplicity up to order four on determinations of the IMF. They found that even in the most extreme cases (100\% stellar multiplicity) the effect on the observed IMF is small, within the uncertainties of most studies. It is only below system masses of 0.5 $M_{\sun}$ that binaries begin to effect the observed luminosity function significantly \cp{kroupa91, kroupa93}, hence the effects of unresolved multiplicity on our results should not be significant. However, to the extent that unresolved stellar multiplicity is present, it would cause us to overestimate the number of high mass stars and bias our results toward a higher $M_U$ or flatter $\alpha$ than the true IMF. 

Interacting binaries may also complicate the interpretation of these results \cp[e.g.]{eldridge09,yang11,li12}. They have been typically been ignored in other studies of resolved stellar populations \cp[e.g.][]{aparicio04,weisz11,lewis15}; as their effects on the observed CMD are dependant on many parameters such as mass transfer and initial binary distribution \cp{larsen11,eldridge17}. 

We model a uniform reddening across the field of view  equal to the foreground  $E(B-V)^G = 0.009$ \cp{jacobs09}  and neglect the effects of internal dust extinction. Although such an approach is common in the literature \cp[e.g.][]{mcconnachie05,williams09a,hillis16} it raises concerns since young stars should be found close to their birth clouds, where the ISM is still dense and clumpy, and young star clusters are likely to have higher levels of internal extinction than old ones \cp{grosbol12}. This would add intrinsic scatter to the CMD which we have not modelled.  We examined the $V-I$ versus $g-V$ two colour diagram and find no detectable net reddening of MS stars to the limit of $E(B-V) \lesssim 0.1$, which is too loose of a constraint to be useful in this study.  We justify our approach by noting that a low amount of internal dust is expected for the outer disc of DDO 154 since it is far removed from the star-forming centre. Furthermore, the dust content in galaxies has been observed to decrease with galactocentric radius \cp{mateos09}, the dust to gas mass ratio is lower in low metallicity systems  \cp{ruyer14} such as DDO 154 ($Z = 0.1 Z_{\sun}$, see Table. \ref{tab:gal_params}). Many dwarf galaxies  have little or no detectable dust \cp{vanzee00};the SINGS \cp{kennicutt03} and KINGFISH \cp{kennicutt11,ruyer14} surveys detect no far infrared dust emission from DDO 154 using \emph{Spitzer} and \emph{Herschel} data. Hence scatter on the CMD due to internal extinction should not be significant.

Overall, neglecting stellar rotation and multiplicity would bias us toward overestimating the number of high mass stars, or equivalently deriving a flatter $\alpha$ or higher $M_U$ than the true IMF. As the best-fitting IMF is already steeper and deficient in high mass stars compared to a universal Kroupa IMF, the true deficiency of high mass stars is likely underestimated by our results.

\section{Star formation in the outer disc} \label{sec:5}
\subsection{Local star formation law}
\begin{figure}
\includegraphics[width=0.45\textwidth]{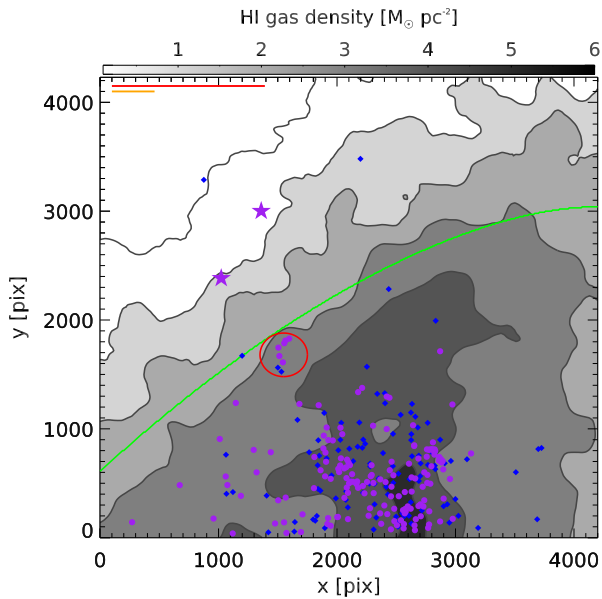}
\caption{The spatial correlation between MS (purple) and BL (blue diamonds) stars and the surface density of \HI\ in the \emph{HST}/ACS field of view. \HI\ contours are at 0.5, 1, 2, 4 and 6 M$_{\sun}$ pc$^{-2}$. The red and orange bars in the top left corner denote the  distances stars in our selection box could travel over the MS lifetime of a 4 M$_{\sun}$ and  26 $M_{\sun}$ star respectively if travelling at the \HI\ velocity dispersion The green contour corresponds to a radius of 6 kpc in the disc plane.  \label{fig:hi_ms}}
\end{figure}

Figure \ref{fig:hi_ms} shows the correlation between the MS stars and the \HI\ surface density ($\Sigma_{\text{\HI}}$) from the  THINGS  survey \cp{walter08}.  We used the natural weighted map (resolution 6$''$  = 117 pc) for better sensitivity to diffuse emission. The MS stars are mostly located in the bottom half of our images, as displayed in Figures \ref{fig:spatial} and \ref{fig:hi_ms}, closest to the optically bright central region of the galaxy (beyond the edge of our images) and where $\Sigma_{\text{\HI}}$ is the highest within our frame (Fig. \ref{fig:uv}). There are very few MS stars beyond $R \approx 6 $ kpc where  $\Sigma_{\text{\HI}} \leq 2 \ M_{\sun}$ pc$^{-2}$.  Even the BL stars which trace the formation of stars with masses down to  $\sim 2.6 \ M_{\sun}$, and hence ages  up to $\sim 420$ Myr ago, appear to be scarce in the outer disc.  The red bar in Fig. \ref{fig:hi_ms} corresponds to the maximum distance a 4 $M_{\sun}$ star could travel  over its MS lifetime travelling at the typical \HI\  velocity dispersion in our \emph{HST}  images of 8.62 km s$^{-1}$.  The two farthest MS stars fall within this tolerance of ASN-1 (marked by a red circle). Hence it is unclear whether these stars were formed in situ at high radii, or have drifted from a different formation site.  The relative quiescence of the outer disc is an interesting contrast to the studies of \ct{gildepaz05} and  \ct{thilker05,thilker07} who find that UV bright star formation is common in extended outer discs. 

To parameterise the SFL we compare $\Sigma_{\text{\HI}}$ to the  SFR surface density ($\Sigma_{\text{SFR}}$) derived   from the surface density of MS stars  ($\Sigma_{\text{MS}}$). We  count the number of MS stars in bins of 0.1 dex in $\Sigma_{\text{\HI}}$   and divide by the  inclination corrected pixel area covered by each bin,  equivalent to the method used by \ct{sanduleak69} and \cta{sarah15}. $\Sigma_{\text{MS}}$ is then converted to $\Sigma_{\text{SFR}}$ assuming a Kroupa  (or best-fit) IMF slope for $m \geq 1\ M_{\sun}$, and the \ct{kroupa01} broken power law for $0.08 \ M_{\sun} \leq m <1 \ M_{\sun}$.  The conversion factors for a Kroupa IMF and a best-fit IMF are 2.2 and 2.36 $(\times 10^{-6}) \ M_{\sun} \ \text{yr}^{-1} \ \text{star}^{-1}$ respectively. 

\begin{figure}
\includegraphics[width=0.45\textwidth]{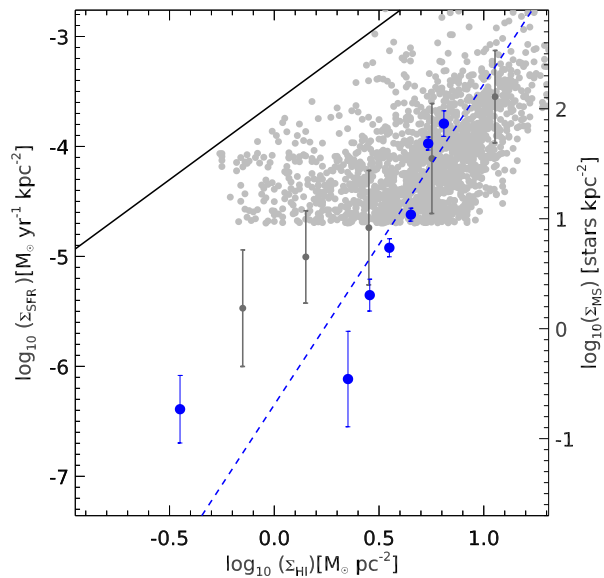}
\caption{The local SFL in the outer disc of DDO 154 derived from the surface density of MS stars assuming a Kroupa IMF shown as blue points with error bars. The blue dashed line represents our linear, error weighted, least squares fit to the data and the black line shows the \ct{kennicutt98} SFL with $N=1.4$ for the bright part of galaxies. The SFL for dwarf galaxies from \ct{bigiel10} is shown as grey points with median values given in darker grey with error bars. \label{fig:sfl}}
\end{figure}

 Fig. \ref{fig:sfl} compares our derived SFL to the standard K-S SFL \cp{kennicutt98} and the SFL derived for dwarf galaxies by \ct{bigiel10}. Here we adopt a Kroupa IMF to be consistent with these and other previous studies.  A linear, error-weighted least squared fit to our data  gives a SFL with slope $N= 2.92 \pm 0.22$ and zero point  $A= 4.47 \pm 0.65 \times 10^{-7} \ M_{{\sun}}\  \text{yr}^{-1} \ \text{kpc}^{-2}$. If we assume our best-fitting IMF the zero point becomes $A=4.82 \pm 0.71  \times 10^{-7} \ M_{{\sun}}\  \text{yr}^{-1} \ \text{kpc}^{-2}$. Star formation appears to become quickly suppressed as $\log_{10} (\Sigma_{\text{\HI}} / M_{\sun} \  \text{pc}^{-2})$ falls below $\sim 0.7$. Our derived power-law slope is typical of the values determined for  dwarf irregular galaxies  by \ct[][at 200 pc and 400pc spatial resolution, compared to our 117pc]{roychowdhury09}, but there are some differences. The total \HI\ mass of DDO 154 is an order of magnitude higher when compared to galaxies with a similar $N$ value, which shows up as a difference in the value of the zero point. Galaxies from \ct{roychowdhury09} such as UGC 8215 ($N$ = 3.1, A = 1.99 $\times 10^{-6} \ M_{{\sun}}\  \text{yr}^{-1} \ \text{kpc}^{-2}$) and DDO 181 ($N$ = 2.6, A = 1.48 $\times 10^{-5} \ M_{{\sun}}\  \text{yr}^{-1} \ \text{kpc}^{-2}$) both have similar $N$ values to  DDO 154, but much higher $A$ values.  It is only when we compare just the  \HI\ mass in our \emph{HST} field of view ($6.1 \times 10^{7} \ M_{\sun}$) to the  total \HI\ mass of these galaxies, that the $N$ values and \HI\ masses show  a correspondence. 
  
   The \ct{bigiel10} data differs from ours as they derive SFR from UV surface brightness and average $\Sigma_{\text{\HI}}$ over 750 pc boxes whereas our SFL is calculated by counting stars within \HI\ isophotes. Nevertheless, the majority of our data points are within the scatter of the \ct{bigiel10} measurements. Interestingly, Fig. \ref{fig:sfl} shows  measuring $\Sigma_{\text{SFR}}$ leads to values significantly lower than those obtained from GALEX. This is also suggested by Fig. \ref{fig:uv} where some MS stars do not have correlated NUV emission.  When compared to the SFL derived in the outskirts of massive galaxies from the The Local Volume \HI\ Survey \cp[LV\HI S,][]{wang17}, their $\Sigma_{\text{SFR}}$ values sit mostly above the median trend of \ct{bigiel10}, whereas ours is consistently below it, reflecting the dormancy of the outer disc of DDO 154.   It is commonly observed that the central regions of dwarf galaxies have a similar SFL to outer discs of spirals \cp[e.g.][]{bigiel10,roychowdhury15}. Our method assumes that the gas and stars have not moved since the birth of the stars. In reality stars may have migrated causing a potential underestimation of $\Sigma_{\text{SFR}}$ in regions of higher $\Sigma_{\text{\HI}}$, and an overestimation where $\Sigma_{\text{\HI}}$ is lower. Potentially biasing $N$ to be flatter than the true SFL. 
 
 \subsection{Comparison to theory}
\begin{figure}
\includegraphics[width=0.45\textwidth]{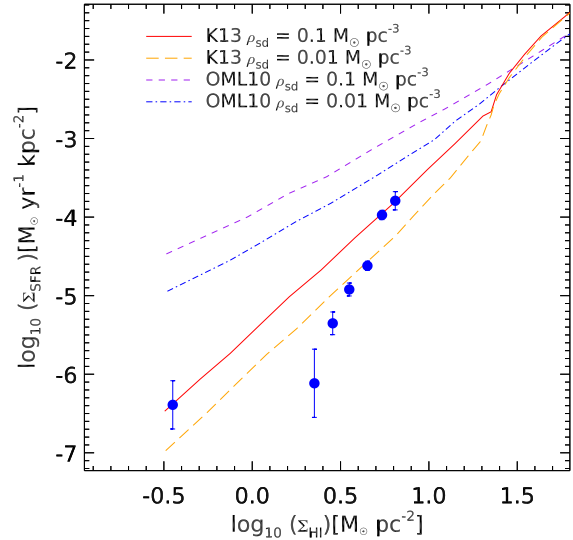}
\caption{ Comparison of the SFL in the outer disc of DDO 154 (the same as Fig. \ref{fig:sfl}) to the theoretical models of  \ct[][ OML10]{ostriker10}  and  \ct[][K13]{krumholz13}. We show models for the combined volume density of stars and dark matter $\rho_{\text{sd}} = 0.1 M_{\sun} \ \text{pc}^{-3}$ and   $\rho_{\text{sd}} = 0.01 M_{\sun} \ \text{pc}^{-3}$ for our adopted metallicity $Z=0.1Z_{\sun}$. \label{fig:sfr_models}}
\end{figure}

\ct{krumholz13}  suggests that differences in the SFL observed between the bright centres of galaxies, and outer discs and dwarf galaxies is due to the processes that govern the conversion of  \HI\ to \HH. He modelled star formation in \HI\ dominated conditions and found results consistent with the observations of \ct{bigiel10}.  In  Fig. \ref{fig:sfr_models} we compare our observed SFL to models  of star formation in \HI\ dominated regions by \ct{ostriker10} and  \ct{krumholz13}. The key parameter in both models is the combined volume density of stars and dark matter, $\rho_{\text{sd}}$, and the gas metallicity in the case of \ct{krumholz13}. Our data favours the \ct{krumholz13} models but is steeper than both,  favouring $\rho_{\text{sd}} =0.1 M_{\sun} \ \text{pc}^{-3}$ at higher $\log_{10}(\Sigma_{\text{\HI}})$ and  $\rho_{\text{sd}} =0.01 M_{\sun} \ \text{pc}^{-3}$ at lower. Overall the data appears to favour the lower density model, so we assume  this density to be the best-fit to the data.

 How does this compare to the stellar densities predicted by the IMF analysis? The expected density of stars depends on the timescale that star formation has persisted. We take the minimum star formation timescale to be the minimum orbital time in the \emph{HST} images ($\sim743$ Myr, at a radius 2.3 kpc) and the maximum being the Hubble time. The volume density of stars depends on the scale height of the disc, for which we choose a typical value of $h\sim 100$pc. Assuming a Kroupa IMF the face on SFR surface density in our optical images is $3.99\times 10^{-6} \ M_{\sun} \ \text{yr}^{-1} \ \text{kpc}^{-2}$, corresponding to  minimum and maximum stellar volume densities for a disc of thickness $2h$ of  $\rho_{\text{s,min}} = 1.48 \times 10^{-5} \ M_{\sun} \ \text{pc}^{-3}$ and $\rho_{\text{s,max}} = 2.73 \times 10^{-4} \ M_{\sun} \ \text{pc}^{-3}$. If our best-fitting IMF is used instead, these values are 1.08 times larger. These  stellar density estimates  are too low to account our SFL results which best match the  \ct{krumholz13} $\rho_{\text{sd}} =0.01\ M_{\sun} \ \text{pc}^{-3}$ theoretical model. Mass modelling of DDO 154 by \ct{carig98} also predicts the density of dark matter alone would also be too low:  $\rho_{\text{d}} =0.015 \ M_{\sun} \ \text{pc}^{-3}$ at 1.4 kpc and $\rho_{\text{d}} =0.001 M_{\sun} \ \text{pc}^{-3}$ at 6.1 kpc. 
 We determine the maximum allowable mass density using the THINGS rotation curve \cp{walter08}  to find the mass enclosed in the radii range covered by our \emph{HST} images (2.3 kpc --- 12.2 kpc). Assuming a uniform density ring of thickness $2h$ we find  $\rho_{max} = 0.061 \ M_{\sun} \ \text{pc}^{-3}$, which is higher than the \ct{krumholz13} $\rho_{\text{sd}} =0.01\ M_{\sun} \ \text{pc}^{-3}$  model.  Hence, these models  work best if the dark matter is somewhere in-between  a spherical distribution, and a flattened disc like system.  This could imply energy dissipation contrary to the standard CDM scenario  \cp[e.g.][]{davis85,peebles17}, or as  we only consider  the two extreme scenarios there could be geometric factors which we are not accounting for. 

\subsection{Total star formation rate} \label{subsec:int_sfl}
We determine the SFR integrated over the disc of DDO 154 to be $3.14 \times 10^{-3} \ M_{\sun} \ \text{yr}^{-1}$ assuming our derived SFL applies for the entire gas disc with an adopted Kroupa IMF.  This is within the uncertainties of the \HA\ based SFR estimates, and slightly lower than the UV+IR SFR  from the literature  (see Table \ref{tab:gal_params}). Adopting our best-fitting IMF increases the SFR by a factor of 1.08 slightly improving the comparison to  previous estimates of the UV+IR SFR. However, these published values cover just the central region of the galaxy which is largely outside our field of view and where star formation would be expected to be more `normal'. If we restrict the integrated SFR to radii greater than the $R_{25} = 3.61$ kpc \cp{vaucouleurs91} we find the  SFR of the outer disc to be $1.20 \times 10^{-3}  \ M_{\sun} \ \text{yr}^{-1}$ assuming a Kroupa IMF. This is 24\% of the UV+IR SFR estimate of the central region of the galaxy, a significant but not dominant contribution to the total SFR of the galaxy. This corresponds to a face on SFR surface density for $R>R_{25}$ of $2.80 \times 10^{-6}  \ M_{\sun} \ \text{yr}^{-1} \ \text{kpc}^{-2}$, which is typical of the intensities observed in the outer discs of spirals \cp[$10^{-6} $---$ 10^{-5} M_{\sun} \ \text{yr}^{-1} \ \text{kpc}^{-2}$, ][]{barnes12} and other dwarf galaxies \cpa[e.g. NGC 2915: $1.8 \times 10^{-5}  M_{\sun} \ \text{yr}^{-1} \ \text{kpc}^{-2},$][]{sarah15}.

\subsection{IGIMF theory} \label{subsec:igimf}
\begin{figure}
\includegraphics[width=0.45\textwidth]{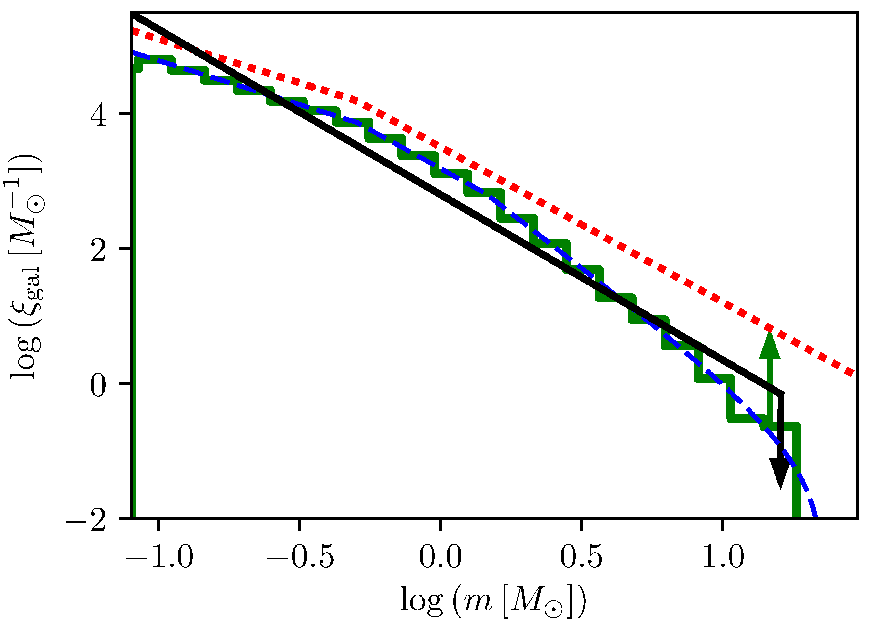}
\caption{Comparison between the best-fit IMF (black), Kroupa IMF (red, offset +0.5 dex), prediction from IGIMF theory (blue, dashed), and the optimally sampled galaxy-wide IMF (green). The upper mass limit from the best-fit IMF $M_U = 16\ M_{\sun}$ is shown by a black downward pointing arrow, and the optimally sampled prediction $m  = 15\ M_{\sun}$ by a green upward arrow.   \label{fig:igimf} }
\end{figure}

So far, we have modelled star formation purely stochastically: without clustering in space or time.
However, the formation of stars appears to be highly correlated, with new stars tending to form in embedded clusters, associations, or groups many of which later dissolve \cp{lada03}.  One approach to account for this is to use the Integrated Galaxy-wide Initial Mass Function (IGIMF) framework \cp{weidner05, weidner06,weidner10, yan17}, whereby all stars are born in clusters adhering to a well behaved IMF with an upper mass limit that varies but is set by the cluster mass, a SFR-dependant single-slope power law cluster mass function \cp{yan17}, and a maximum cluster mass set by the SFR. When integrated over all the star-forming clusters in a galaxy that is weakly forming stars the IGIMF is top light compared to the universal IMF (it is top heavy at high star formation rates).
In Fig. \ref{fig:igimf} we compare the best-fit IMF we derive for the outskirts of DDO154 ($M_U = 16 \  M_{\sun} $, $\alpha = -2.45$), with the distribution of stellar masses we expect from IGIMF theory following \ct{yan17}. We use our adopted metallicity ($Z=0.1\ Z_{\sun}$) and a SFR of $4.65 \times 10^{-4} \ M_{\sun}\ \text{yr}^{-1}$, the total SFR in our \emph{HST}/ACS field of view assuming a best-fit IMF.  For comparison we also display a standard Kroupa \cp{kroupa01} IMF (with an offset so as not to clutter the figure).  We see that our best-fit IMF is very similar to that predicted by IGIMF theory.  The low SFR of DDO154 is thus similar to the low density outskirts of the Orion nebula which appear deficient in O and B stars \cp{megeath16}. 

\subsection{Stability of the outer disc}
\begin{figure}
\includegraphics[width=0.45\textwidth]{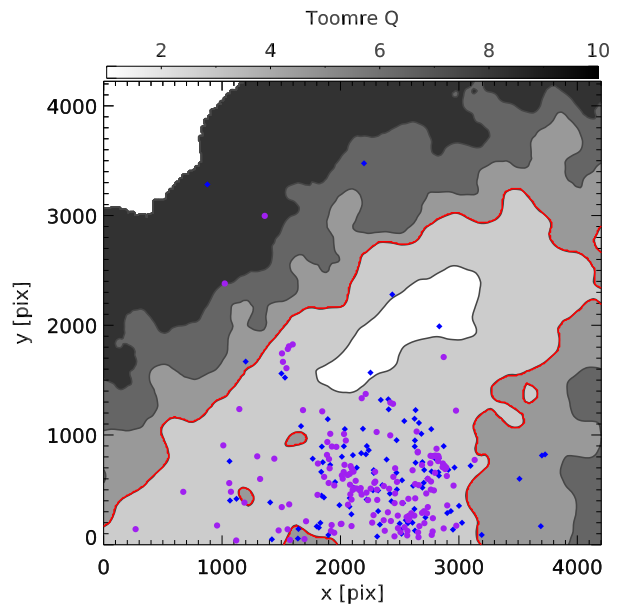}
\caption{Contour plot of the Toomre $Q$ parameter across the \emph{HST}/ACS field of view with MS (purple) and BL (blue diamonds) shown for spatial comparison. The contours are spaced at $Q=$ 2, 4, 6, and 10 with the scale given by the colour bar at the top. The red contour corresponds to $Q = 4$, and the white region in the top left is where the data ends. \label{fig:toomre}}
\end{figure}

\begin{figure}
\includegraphics[width=0.4\textwidth]{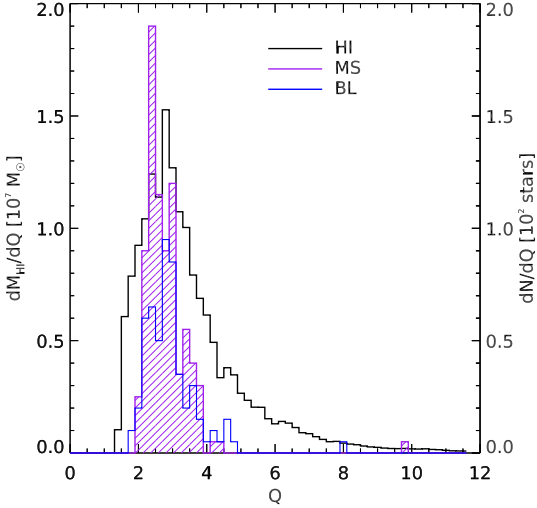}
\caption{Histogram comparing  the distribution of \HI\ mass (black, left axis) in the \emph{HST} field of view and the number of MS (purple, shaded) and BL (blue) stars (right axis) as a function of the $Q$ parameter.  \label{fig:qhist}}
\end{figure}

\begin{figure}
\includegraphics[width=0.4\textwidth]{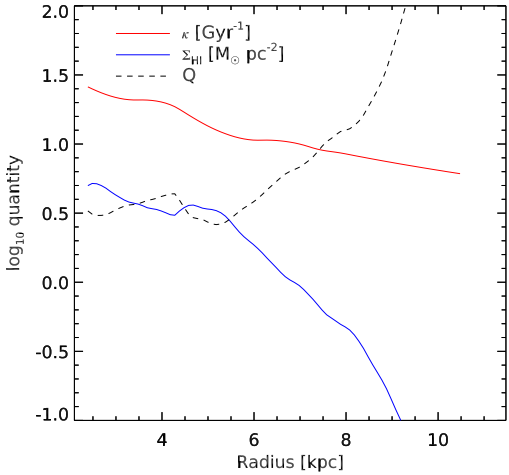}
\caption{Radially averaged profiles of $\kappa$, $Q$ and $\Sigma_{\text{\HI}}$ in the \emph{HST} field of view. The rapid increase in $Q$ beyond $R=6$ kpc coincides with steeper decline in $\Sigma_{\text{\HI}}$, while $\kappa$ falls steadily.   \label{fig:rad_q}}
\end{figure}

 Insight into the dynamics of the dormant and star-forming parts of the outer disc can be provided by the \HI\ data.   Using the THINGS rotation curve  \cp{walter08}  and adopting a flat velocity dispersion of $\sigma_g = 8.62$ km s$^{-1}$   we create a spatial map of the Toomre $Q$ parameter of the \HI\ in our optical images. We transform and re-grid the maps to the field of view and resolution of our optical images, and  calculate radii in the disc plane assuming the tilted ring model  of \ct{deblok08}.  The  \ct{toomre64}  $Q$ parameter is given by
 \begin{equation}
Q\ =\ \frac{\sigma_{g} \kappa}{\pi G \Sigma_{\text{gas}}},
\end{equation}
where $\sigma_g$ is the gas velocity dispersion, G is the gravity constant, and  $\kappa$ is the epicyclic frequency given by
\begin{equation}
\kappa\ =\ 1.41 \frac{V_{c}(R)}{R} \sqrt{1+\frac{d\log V_{c}(R)}{d\log R}}.
\end{equation}
Adopting  $\Sigma_\text{gas} = \Sigma_{\text{\HI}}$,   Fig. \ref{fig:toomre} shows the distribution of the $Q$ parameter in our \emph{HST} images.  Almost all of the MS and BL stars appear to be in regions where $Q \leq 4$. We compare histograms of  \HI\ mass and star counts as a function of $Q$  in Fig. \ref{fig:qhist}. There is a strong peak in \HI\ mass and star counts at $Q\approx 2$ to 3, and a clearly extended tail of high $Q$ gas corresponding to the extended outer disc (the dark regions in Fig. \ref{fig:toomre}). Quantitatively,  28\% of the \HI\ mass  in our \emph{HST}/ACS images has stability parameter $Q\geq4$ while only 2.5\% of the MS stars and 9.3\% of the BL stars exist in regions with this stability and higher. In Fig. \ref{fig:rad_q} we compare radially averaged profiles of $Q$, $\kappa$ and $\Sigma_{\text{\HI}}$. The rise in $Q$ beyond $R\approx 6$ kpc can be seen to be driven by the rapid decrease in $\Sigma_{\text{\HI}}$ which is proceeding faster than the decrease in $\kappa$. This is consistent with \ct{meurer13}, who observe $Q$ transition from flat to rising at high radii in disc galaxies.   

The high stability of the outer disc may be causing the absence of young stars. Star formation traced by \HA\ emission is typically truncated when the stability parameter Q exceeds a threshold, corresponding to $Q \approx 2.1$ \cp{kennicutt89, martin01}. This  coincides with the peak $Q$ of the young stars in Fig. \ref{fig:qhist} implying that star formation is truncated where the disc is stable; hence our results are consistent with \ct{kennicutt98} and \ct{martin01}. However, this result is in contrast to the discovery that outer discs often exhibit UV bright star formation even where \HA\ emission is weak or absent  \cp{gildepaz05,thilker05,thilker07}.  The outer disc of DDO154 is nearly dormant according to all available star formation tracers, but may contain star formation below our detection limit.

\section{Conclusions} \label{sec:6}
Stellar populations uniquely determine the vast majority of the observed properties of galaxies such as metallicity, surface brightness, and emission line strengths. While the stellar initial mass function (IMF) is commonly assumed to be invariant regardless of the star-forming environment, there is strong evidence of real variations of the properties of its upper mass end \cp{hoversten08,meurer09,lee09,dabringhausen10,gunawardhana11,marks12, romano17,schneider18}. We have placed constraints on the high mass slope of the IMF using the Main Sequence Luminosity Function extracted from an optical colour magnitude diagram (CMD) of the resolved population of stars in the extended \HI\  outer disc of the dwarf irregular galaxy DDO 154.

Our main results are as follows: The IMF  required to match the observed population of main sequence (MS) stars is deficient in high mass stars compared to a standard Kroupa IMF \cp{kroupa01}. The Kroupa IMF is ruled out to a $p=0.088$ significance level using a KS test between the observed MS stars and our simulated stellar populations. We find that the star formation rate (SFR) in the outer disc is consistent with observations of other dwarf galaxies, where star formation is suppressed below the standard Kennicutt-Schmidt law observed in the bright portions of galaxies \cp{kennicutt89, kennicutt98,martin01, bigiel08}. Our best-fit IMF agrees with the predictions of IGIMF theory, thus star formation in this outer disc is similar to the low density regions of the Orion nebula \cp{megeath16}.   Deeper observations of extended discs would provide insight  to whether low mass stars are forming, as these observations point to an IMF deficient in the highest mass stars.

A stand out feature of previous studies is that star formation in the outer discs of galaxies is common \cp[e.g.][]{gildepaz05,thilker05,thilker07} and appears to follow the \HI\  (\citealt{cuillandre01, pflamm08}; \cta{sarah15}). In this work, we find  star formation is absent in much of outer disc of DDO 154, at least in terms of stars of mass $m>4 \ M_{\sun}$. The high stability parameter $Q>4$ of the  extended \HI\ disc demonstrates that some galaxies can maintain a large reservoir of ISM which is dormant in high mass star formation. It is not known how common these dormant discs are.

\section*{Acknowledgements}
We thank the anonymous referee for their helpful comments that improved this paper. AW, GRM, KP and TJ acknowledge financial support through the DAAD (The Australian - Germany Joint Research Co-operation Scheme) throughout the course of this work. PK and TJ thank the DAAD for supporting this project with travel grant 57212729, "Galaxy formation with a variable stellar initial mass function". TJ was supported by the University of Bonn and by Charles University through the grant SVV-260441. 

This project was initiated as part of the Advanced Camera for Surveys (ACS) Instrument Definition Team effort. ACS was developed under NASA contract NAS 5-32865, and this research has been supported by NASA grant NAG5-7697 and by an equipment grant from Sun Microsystems, Inc. Some/all of the HST and POSS data presented in this paper were obtained from the Mikulski Archive for Space Telescopes (MAST). STScI is operated by the Association of Universities for Research in Astronomy, Inc., under NASA contract NAS5-26555. Support for MAST for non-HST data is provided by the NASA Office of Space Science via grant NNX09AF08G and by other grants and contracts. 




\bibliographystyle{mnras}
\bibliography{paperbib.bib}



\bsp	
\label{lastpage}
\end{document}